\documentclass[twocolumn,showpacs,preprintnumbers,floatfix,pre,a4paper]{revtex4}
\usepackage{amssymb,amsmath}
\usepackage{graphicx}
\usepackage{bm}

\begin{document}
\title{Rolling/Slipping Motion of Euler's Disk}
\author{H.Caps} \author{S.Dorbolo} \author{S.Ponte} \author{H.Croisier} \author{N.Vandewalle}           
\affiliation{GRASP, Department of Physics B5, University of Li\`ege, B-4000 Li\`ege, Belgium.}

\date{Received: \today / Revised version: date}

\begin{abstract} 
We present an experimental study of the motion of a circular disk spun onto a table. With the help of a high speed video system, the temporal evolutions of (i) the inclination angle $\alpha$, (ii) the angular velocity $\omega$ and (iii) the precession rate $\Omega$ are studied. The influence of the mass of the disk and the friction between the disk and the supporting surface are considered. 
The inclination angle $\alpha$ and the angular velocity are observed
to decrease according to a power law. We also show that the precession
rate $\Omega$ diverges as the disk stops. Exponents are measured very
near the collapse as well as on long range times. Collapsing times
have been also measured. The results are compared with previous theoretical and experimental works. The major source of energy dissipation is found to be the slipping of the disk on the plane.
\end{abstract}



\maketitle

\section{Introduction}
A common straightforward experiment is the following. A coin spun onto
a table rotates with an increasing precession rate and a decreasing
inclination angle until it abruptly stops. This complex motion is known as the Euler's disk problem.

Within the classical formalism of mechanics, the disk should spin {\em ad infinitum}. Moffat's work \cite{moffatt1,nature} emphasized the necessity of considering energy dissipation in order to avoid such forever persisting motion. Many theoretical and numerical studies \cite{physd,kessler,pre2,mcdonald} have been devoted to the finding of the major dissipation process leading to the stop of the disk. However, only a few experimental result can be found in the scientific literature \cite{pre1,nature}. 

\begin{figure}[htb]
\begin{center}
\includegraphics[width=7cm]{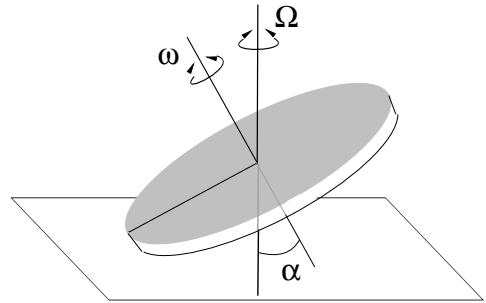}
\caption{Sketch of the disk and its dynamical parameters: the inclination angle $\alpha$, the precession speed $\Omega$ and the angular velocity $\omega$.}\label{param}
\end{center}
\end{figure}

For a disk of radius $R$, mass $m$, and an inertia momentum $I$, the total energy reads
\begin{equation}\label{eqenergy}
E=mgR\sin(\alpha)+\frac{1}{2}I\Omega^2 \sin^2(\alpha),
\end{equation} 
where $\alpha$ is the inclination angle with respect to the horizontal and $\Omega$ is the precession rate. A sketch of the disk is illustrated in Figure~\ref{param}. Parameters $\alpha$, $\Omega$ and $\omega$ are emphasized.

In order to meet observations, one has to consider that the energy $E$
is dissipated with a rate $\phi$. The goal is thus to find the major
mechanism responsible for energy dissipation $\phi$.

Considering that the energy is dissipated through a velocity-dependent process, the temporal evolutions of $\alpha$, $\Omega$ and $\omega$ can be described in a general way with 
\begin{equation}\label{eqalpha}
\alpha\sim (t_0-t)^{n_\alpha},
\end{equation}
\begin{equation}\label{eqOmega}
\Omega\sim \left(\frac{1}{t_0-t}\right)^{n_\Omega}
\end{equation}
and
\begin{equation}\label{eqomega}
\omega\sim (t_0-t)^{n_\omega},
\end{equation}
where $n_\alpha$, $n_\Omega$ and $n_\omega$ are the dynamical
exponents to be determined \cite{mcdonald}. Depending on the
dissipation process, different values can be found for these
exponents. The prefactors of those laws, and hence collapsing times $t_0$,
also depend on the dissipation mechanism. Moreover, it should be
noticed that the power laws may be valid only in a limited range of
inclination angles since the major dissipation mechanism may depend on
the dynamical parameters $\alpha$ and $\Omega$ values.

Nowadays, no consensus has been found by the physical
community. Other experiments are thus required in order to emphasize the major dissipation process leading to the brutal stop of the disk.

In the next section, we present the
experimental setup and procedures. In Section III, we report the
experimental results concerning the temporal evolution of the
inclination angle, the precession frequency and the rotation velocity
as well as times to collapse. We discuss the results in Section IV. A summary of our findings is eventually given in Section V.

\section{Experimental procedures}

The experiment consists in spinning a disk by hand onto a table and to observe the evolution of the three different dynamical parameters with the help of a high speed video system. 

We used three different coins: (i) a stainless disk having a diameter  $D_2=75\pm 0.01$mm, a thickness $e_2=10\pm 0.01$mm and a mass $m_2=353.4\pm 0.1$g ; (ii) an alumna disk with a diameter $D_1=75\pm 0.01$mm, a thickness $e_1=9.9\pm 0.01$mm and a mass $m_1=123.9\pm 0.1$g; (iii) an alumna torus with an internal diameter $D_{T_i}=63.65\pm 0.01$mm, an external diameter  $D_{T_e}=78\pm 0.01$mm, a thickness $e_T=10\pm 0.02$mm and a mass $m_T=124.5\pm 0.1$g. 

The used surfaces on which the disks spin were $2$ mm thickness
plates of alumna, glass and rough plastic sheets. All these plates
where supported by the same table in order to change only the surface
roughness. Starting from the lowest friction coefficient to the
uppest, one finds the glass, the alumna and the rough plastic surface.

\

Herebelow, we describe the experimental procedures we used for the
data acquisitions. Note that despite the fact that the disks are spun
by hand, our results are quite reproducible. This has been also mentioned in \cite{pre1}.

\subsection{Inclination angle $\alpha$}

\begin{figure}[htb]
\begin{center}
\includegraphics[width=7cm]{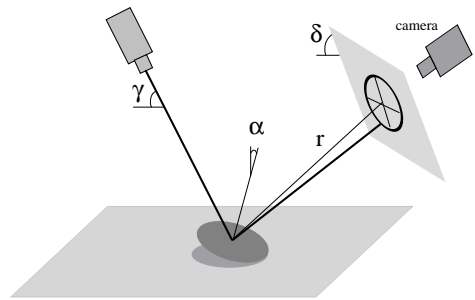}
\caption{Experimental setup for the inclination angle measurement. A laser beam is reflected by the disk and creates an `ellipse' on a screen situated on the other side. A high speed video camera records images of the `ellipse'.}\label{ellipse}
\end{center}
\end{figure}

The inclination angle measurement is based on the reflectivity of the disks. When a laser beam is directed to a disk with a finite angle $\gamma$ relative to the horizontal plane, one can observe an ``ellipse'' formed by the laser beam on a screen situated on the opposite side. This ellipse is due to the motion of the disk, which changes the incident angle of the laser beam on the disk. In fact, a circle should be observed on the screen. However, the angle between the laser beam and the horizontal plane induces a deformation of that circle into an ellipse. In order to minimize this effect, the laser beam is placed as vertically as possible ($\gamma\approx 80^\circ$ in practice); its position being limited by the screen. A sketch of the setup is illustrated in Figure~\ref{ellipse}. 

\begin{figure}[htb]
\begin{center}
\includegraphics[width=5cm]{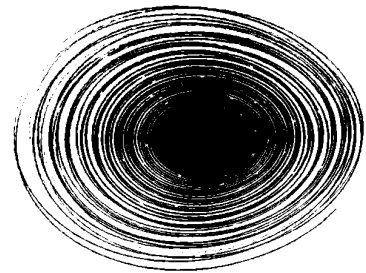}
\caption{Path of the laser beam during the disk rotation. The size of the ellipse decreases with time.}\label{spirale}
\end{center}
\end{figure}

Since the inclination angle $\alpha$ of the disk decreases with time, the size of the ellipse accordingly decreases. Measuring the size of the ellipse gives us the inclination angle $\alpha$ of the disk as a function of time. In Figure~\ref{spirale}, we present the whole path of the laser beam during the last seconds of the disk motion. One can observe a spiral pattern due to the decrease of $\alpha$ with time. The black circle in the center of the picture is an artifact due to the width of the trajectory on the picture. Experimentally, the high speed video camera allows us to distinguish the trajectories until $0.0005\, s$ before the disk motion stops. However, direct observations from the side of the disks showed that a recording rate of 500 frames per second is large enough for an accurate determination of $\alpha$. 

Geometrical arguments give us the relation between the size of the ellipse and the inclination angle $\alpha$. The main axis length $A$ of the ellipse and the inclination angle $\alpha$ are correlated through 
\begin{equation}\label{A} 
 A=\frac{2r\sin(2\alpha)}{\sin(\gamma+\delta)-\frac{\cos^2(\gamma+\delta)}{\sin(\gamma+\delta)}\tan^2(2\alpha)}
\end{equation} 
where $\delta$ is the angle between the screen and the horizontal plane and $r$ is the distance from the center of the disk to the center of the ellipse. Typically, $r=0.3$\, m. This relation is invertible for determining $\alpha$ from the knowledge of $A$, but has a rather complicated formulation. The minor axis case is more simple. Calculations lead to the following relation between the vertical axis length $B$ and the inclination angle
\begin{equation}\label{B} 
\alpha=\frac{1}{2}\arccos\left(\frac{1-2cos(2\beta)+\cos\left(2\arcsin\left(\frac{B}{2r}\right)\right)}{4\sin^2(\beta)}\right)
\end{equation}

For each measured value of $A$ and $B$, we have calculated $\alpha$ from Eqs.\ref{A} and \ref{B} and we have averaged the values.

\subsection{Precession rate $\Omega$}

\begin{figure}[htb]
\begin{center}
\includegraphics[width=7cm]{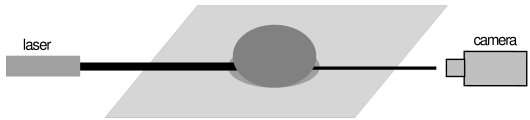}
\caption{Sketch of the experimental setup for the measurement of the precession rate $\Omega$. A laser beam skims the supporting surface and is intercepted by the disk twice a precession. A high speed video camera records the intensity of the transmitted beam.}\label{d_rasemotte}
\end{center}
\end{figure}

In order to measure the precession rate $\Omega$, we have proceeded as
follows. A laser beam has been horizontally placed in front of the
table in such a way that the beam skims the table [see Figure~\ref{d_rasemotte}]. During the rotating motion of the disk, the laser beam is intercepted, leading to a fast decrease of the light intensity transmitted on the other side of the table. Such a light extinction occurs twice a rotation of the disk. The transmitted light is observed with a high speed video camera situated at the same height as the laser but on the opposite side of the plate. The images are then transfered to a computer for image analysis of the intensity of the light spot created by the laser beam. The images are recorded at a frame rate of 500 frames per second, allowing us a high precision for the spot extinction detection.

\begin{figure}[htb]
\begin{center}
\includegraphics[width=6cm]{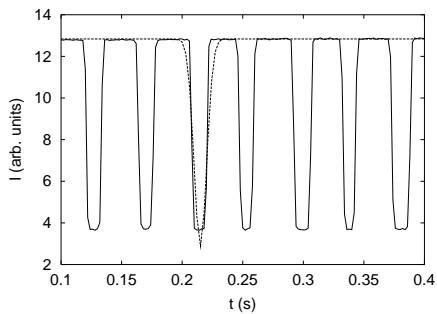}
\caption{Typical curve of the temporal evolution of the intensity $I$ of the laser beam. A fit using a Gaussian curve (dashed curve) is also illustrated.}\label{example}
\end{center}
\end{figure}

Figure~\ref{example} presents a typical curve for the temporal evolution of the intensity $I$ of the laser spot in arbitrary units. Each sharp peak corresponds to an interception of the laser beam by the disk. The peaks are alternatively due to the front and to the back of the disk. The peaks are detected by fitting a Gaussian curve on each point of the signal. In so doing, times corresponding to successive minima of the intensity are recorded. This method allows a precision of $0.002\ s$.

\subsection{Angular velocity $\omega$}

\begin{figure}[htb]
\begin{center}
\includegraphics[width=7cm]{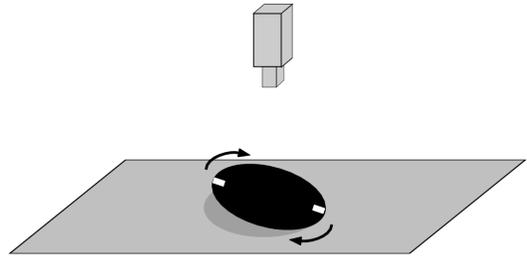}
\caption{Sketch of the experimental setup for determining the angular velocity $\omega$. Two diametrically white dots are glued on the disk. Their motions are recorded with the help of a high speed video camera, from the top.}\label{d_angulaire}
\end{center}
\end{figure}

We have measured the angular velocity $\omega$ of the disk by recording the motion of two diametrically opposed white marks situated on the top face of the disk. The experimental setup is illustrated in Figure~\ref{d_angulaire}. The angular position, and then the angular velocity $\omega$, are calculated from the displacement of both dots between two consecutive frames. The images have been recorded at a rate of $125$ frames per second. This has been found to be sufficient for detecting the stop of the disk. Top view recordings also allowed us to measure the motion of the center of mass of the disk.

\section{Experimental results}

In this section, we present the results of the experiments performed
with three `disks' on the three different surfaces. This corresponds
to nine different experimental situations. The measurements have been
performed a large number of times for each experimental condition. 

We will first consider times to collapse $t_0$ and long time behaviour of
the dynamical parameters. Then, the last stage of the disk's motion
will be further analysed.

\subsection{Time to collapse $t_0$}

\begin{figure}[htb]
\begin{center}
\includegraphics[width=7cm]{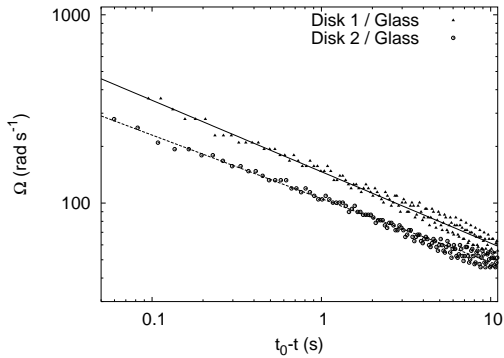}
\caption{Temporal evolution of the precession rate $\Omega$ over an
  entire run in a log-log scale. Two cases are illustrated, as well as fits using
  Eq.(\ref{eqomega}).}\label{longomega}
\end{center}
\end{figure}

Figures \ref{longomega} presents two typical evolutions
of the precession rate $\Omega$ and the inclination angle $\alpha$ as
a function of time $t$. Black lines correspond to fits using Eq.(\ref{eqomega}).

At the beginning of the disk's motion, a larger dispersion of the
measurement is observed; due to the way the disk was set into
motion. The center of mass of the disk may, for exemple, own some
small linear momentum. However, the motion considered in the
introduction is rapidly reached and we have observed that the early
stage of motion has not significant effect on the results presented
hereafter. This was also mentionned in \cite{pre1}.  

One can seen that the power law behavior previoulsy proposed is
consistent with observation over two decades in times. This means
that, as close as we can approach the collapse, the major mechanism of
energy dissipation remains the same whatever its nature.

If the friction between the disk and the surface is increased, the
time to collapse $t_0$ decreases. For the alumna disk (disk 1) on a
glass surface, we have $t_0=12.389\pm 0.007$ while $t_0=3.393\pm
0.168$ for the same disk on the alumna surface (similar results are
obtained for other experimental conditions). Since the geometry of
the system is the same in both cases, the energy dissipation may not come
from the air drag. Friction seems take the main source of dissipation.

\subsection{Inclination angle $\alpha$}

\begin{figure}[htb]
\begin{center}
\includegraphics[width=7cm]{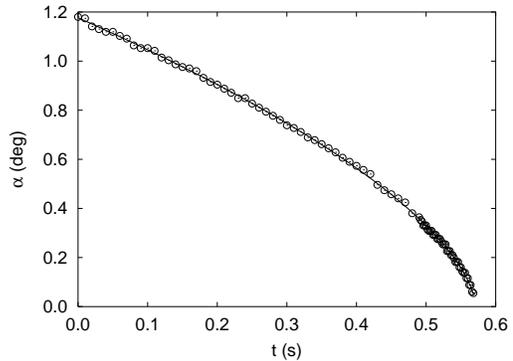}
\caption{Temporal evolution of the inclination angle $\alpha$ for the stainless steel disk (disk 1) case on the alumna surface. A fit using Eq.(\ref{eqalpha}) is also illustrated.}\label{alpha}
\end{center}
\end{figure}

Figure~\ref{alpha} presents a typical curve of the inclination angle $\alpha$ as a function of time $t$. The angle is observed to decrease and to vanish quite abruptly at the time $t_0=0.571\, s$. A fit using Eq.(\ref{eqalpha}) emphasizes the power law behavior. 

The values of the exponent $n_\alpha$ are presented in Table~\ref{tabalpha}. We have observed that despite the error bars, $n_\alpha$ seems to increase when the static friction coefficient $\mu_s$ increases. This tends to show that the friction at the rolling contact point of the disk may play an important role in the dissipation process of the disk energy. One should also note the small differences between the values obtained for the three `disks'. Particularly, the torus behaves in nearly the same way as the two disks. This result suggests that the air layer under the disk is not a major source for the energy dissipation. Another observation is that $n_\alpha$ does not depend on the mass of the spinning coin, in agreement with the equations of motion derived in \cite{mcdonald}.  

\begin{center}
\begin{tabular}{l c c c}
\hline\hline
\bm{$n_\alpha$}  & \quad glass\quad  & \quad alumna\quad  & \quad plastic\quad \\
\hline
disk 1 & \quad0.51 $\pm$ 0.01\quad & \quad0.60 $\pm$ 0.01\quad & \quad0.56 $\pm$ 0.03\quad  \\
disk 2 & \quad0.49 $\pm$ 0.01\quad & \quad0.49 $\pm$ 0.05\quad & \quad0.59 $\pm$ 0.02\quad  \\
torus  & \quad0.52 $\pm$ 0.02\quad & \quad0.56 $\pm$ 0.03\quad & \quad0.66 $\pm$ 0.03\quad  \\
\hline\hline
\end{tabular}
\end{center}
\begin{table}[h]
\caption{Values of the exponent $n_\alpha$ of Eq.(\ref{eqalpha}) for the three `disks' on the different surfaces.}\label{tabalpha}
\end{table}

\subsection{Precession rate $\Omega$}

\begin{figure}[htb]
\begin{center}
\includegraphics[width=7cm]{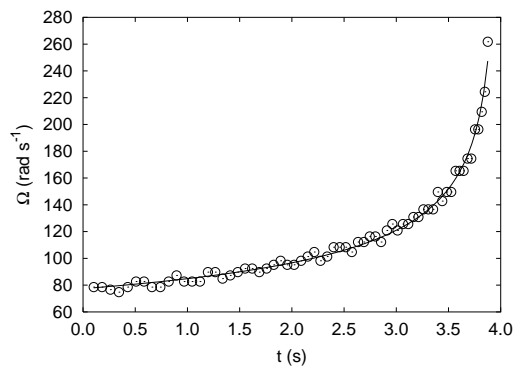}
\caption{Temporal evolution of the precession rate $\Omega$ for the alumna disk case on the alumna surface. A fit using Eq.(\ref{eqOmega}) is also illustrated.}\label{Omega}
\end{center}
\end{figure}

The precession rate $\Omega$ is observed to diverge after a finite
time, as illustrated in Figure~\ref{Omega}. The divergence
(Eq.(\ref{eqOmega})) is illustrated (continuous curve of
Figure~\ref{Omega}). Note that around the black curve, small oscillations of $\Omega$ are observed. They were already reported in other measurements \cite{physd}.

Adjusting the data with the power law Eq.(\ref{eqOmega}), we have found different $n_\Omega$ values for different experimental conditions [see Table~\ref{tabOmega}]. It is observed that the precession rate diverges as rapidly as the static friction increases. Once again, the torus and the disks behave in similar ways, in the sense that the values of $n_\Omega$ are not so different for these different coins. This also suggests that the air drag is not a major source for the energy dissipation.

\begin{center}
\begin{tabular}{l c c c}
\hline\hline
\bm{$n_\Omega$}  & \quad glass\quad & \quad alumna\quad & \quad plastic\quad\\
\hline
disk 1  &\quad 0.39 $\pm$ 0.03\quad  &\quad 0.34 $\pm$ 0.01\quad  &\quad 0.39 $\pm$ 0.03\quad \\
disk 2  &\quad 0.32 $\pm$ 0.01\quad  &\quad 0.30 $\pm$ 0.01\quad  &\quad 0.34 $\pm$ 0.01\quad \\
torus   &\quad 0.30 $\pm$ 0.03\quad  &\quad 0.31 $\pm$ 0.02\quad  &\quad 0.35 $\pm$ 0.02\quad  \\
\hline\hline
\end{tabular}
\end{center}
\begin{table}[h]
\caption{Values of the exponent $n_\Omega$ of Eq.(\ref{eqOmega}) for the three `disks' on the three different surfaces.}\label{tabOmega}
\end{table}

\subsection{Angular velocity $\omega$}

\begin{figure}[htb]
\begin{center}
\includegraphics[width=7cm]{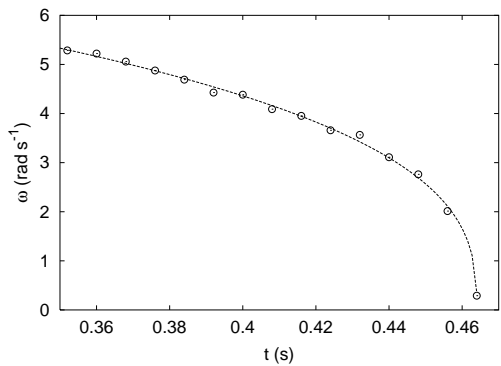}
\caption{Temporal evolution of the angular velocity $\omega$ for the stainless steel disk case on the glass surface. A fit using Eq.(\ref{eqomega}) is also plotted.}\label{omega}
\end{center}
\end{figure}

We have measured the angular velocity of the different coins and reported the values of $n_\omega$ in Table~\ref{tabomega}. A typical curve, as well as a fit using Eq.(\ref{eqomega}) are reported in Figure~\ref{omega}.

As observed for both $\alpha$ and $\Omega$, an increase in the friction between the disk and the surface causes a faster energy dissipation. Here, we observe that the different coins give different values of the exponent $n_\omega$, contrary to previous observations.

\begin{center}
\begin{tabular}{l c c c}
\hline\hline
\bm{$n_\omega$}  &\quad glass\quad &\quad alumna\quad &\quad plastic\quad\\
\hline
disk 1 &\quad 0.38$\pm$ 0.07\quad &\quad 0.46$\pm$ 0.02\quad &\quad 0.41$\pm$ 0.01\quad \\
disk 2 &\quad 0.44$\pm$ 0.01\quad &\quad 0.53$\pm$ 0.06\quad &\quad 0.51$\pm$ 0.02\quad \\
torus  &\quad 0.42$\pm$ 0.01\quad &\quad 0.42$\pm$ 0.01\quad &\quad 0.65$\pm$ 0.03\quad \\
\hline\hline
\end{tabular}
\end{center}
\begin{table}[h]
\caption{Values of the exponent $n_\omega$ for the three `disks' on the different surfaces.}\label{tabomega}
\end{table}

\section{Discussion}

Assuming that the main dissipation mechanism that leads to the
singular behavior is the air drag, Moffatt concluded that $n_\alpha$
should be $1/3$, and $n_\Omega=1/6$. Bildsten pointed out a mistake in
the Moffatt's calculations and found the values $n_\alpha=4/9$ and
$n_\Omega=2/9$. These theoretical previsions are a too low to explain the experimental results, suggesting that air drag does not play an important role in the energy dissipation, as also shown by torus experiments. Experiments performed in vacuum conditions \cite{nature} have also emphasized the small effects of viscous drag.

From the values of both exponents $n_\alpha$ and $n_\Omega$, we thus
see that the rolling friction seems to play an important role in the
energy dissipation mechanism. Indeed, one would expect \cite{pre1}
$n_\Omega=1/3$ in the case of rolling without slipping, what is quite
close to our values. 

This result is also supported by plots of the dissipation rate $\phi$ of
energy during the experiments. We have calculated the total energy $E$
of the disks (Eq.(\ref{eqenergy})) for each measured values of
$\alpha$ and $\Omega$. Then we have computed the dissipated rate of
energy $\phi$ by finite differencing of the energy $E$. 

\begin{figure}[htb]
\begin{center}
\includegraphics[width=7cm]{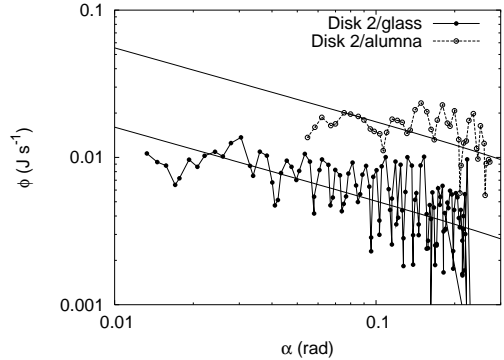}
\caption{Disipation rate of energy $\phi$ as a function of the inclination angle $\alpha$. Two cases are illustrated.}\label{phi}
\end{center}
\end{figure}

In Figure~\ref{phi}, one can find a typical evolution of the dissipation
rate $\phi$ as a function of the inclination angle $\alpha$ for the stainless disk
(disk 2) on both glass and alumna surfaces. Indeed, the disk is osberved to
dissipate more energy by time unit on the surface with the largest
friction. On the same figure, we have plotted the curves corresponding
to a dissipation rate due to friction \cite{pre1}

\begin{equation}
\phi_{frict}\sim\cos(\alpha)\,\Omega
\end{equation}.

The agreement between experimental and theoretical results emphasize
the role played by friction in the energy dissipation. 

Moreover, one should note that a proportionality between the
precession rate $\Omega$ of the disk and its angular velocity $\omega$
is expected \cite{moffatt1,mcdonald}. This is not so obvious here, even if the data are quite dispersed [see Tables~\ref{tabOmega} and \ref{tabomega}]. Moreover, for a considered surface, the angular velocity exponent $n_\omega$ depends on the disk, and therefore on the static friction.

These observations lead us to believe that the disk may slip while it is rotating. The influence of the slipping of the disk was also suggested in \cite{nature}, but from a theoretical point of view. This is in contradiction with \cite{petrie} where the energy dissipation by slipping is ruled out. However, these preliminary experiments were performed during the {\it early} stage of the disk rotation. Indeed, only one experimental condition was measured and moreover at a low frame recording rate. 

\begin{figure}[ht]
\begin{center}
\includegraphics[width=5cm]{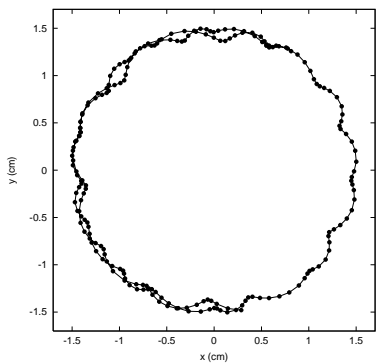}
\caption{Path of the center of mass of the disk 1 on the alumna surface.}\label{cm}
\end{center}
\end{figure}

In order to emphasize the slipping or not of the disk, we have recorded the motion of the center of mass of the disk with a high precision ($250$ fps, $1024\,\times 1024$pixels images). Figure~\ref{cm} illustrates the trajectory of the center of mass within one and a half angular revolution. One can see that the center of mass moves along a circular path. As noticed, the center of this circular path has a tendency to move the most at the early stage of the motion than at the end. In the case of the circular ring, this `stabilization' is less important. We suppose that the large surface of the disks help them to get this stabilization. It will be interesting to check whether this mechanism always occurs in vacuum condition, in order to determine the role of the air in this phenomenon.

Another important observation is that the center mass of the disk exhibits small quasi-periodic excursions around the circular trajectory. These oscillations are not exactly periodic but are quite frequent, especially during the last angular rotations of the disk.   

These small oscillations in the circular trajectory of the center of mass of the disk are due to the slipping of the contact point of the disk along the surface. The resolution of our setup does not allows us to determine wheter there is some kind of `stick-slip' of this point on the surface. However, these slips are responsible for the energy dissipation leading to the final stop of the disk.

\section{Summary}
We have performed new experiments for studying the Euler's disk
motion. The divergence of the precession rate as the inclination angle
and the angular velocity vanish have been observed after a finite time. All
these parameters follow power law behaviors. The time to collaspe is
observed to decrease with the friction. 

Contrary to previous theoretical works, the air is found to be a minor source of energy dissipation. On the contrary, the major energy dissipation process is found to be the rolling/slipping of the disk on the supporting surface. 

\begin{acknowledgements}
HC is financially supported by the FRIA (Brussels, Belgium). SD acknowledges the FNRS (Brussels, Belgium) for financial support. This work is also supported through the ARC contract n$^\circ$02/07-293. 
\end{acknowledgements}

\end{document}